\documentclass[conference,hidelinks]{IEEEtran}
\IEEEoverridecommandlockouts
\usepackage{cite}
\usepackage{amsmath,amssymb,amsfonts}
\usepackage{algorithmic}
\usepackage{graphicx}
\usepackage{siunitx}
\usepackage[T1]{fontenc}
\usepackage{listings}
\usepackage{textcomp}
\usepackage{xcolor}
\usepackage{tikz}
\usepackage{color}
\usepackage{fancyvrb}
\usepackage{orcidlink}
\usepackage{cleveref}
\usepackage{glossaries}
\usepackage[position=b]{subcaption}
\usepackage[a4paper, total={184mm,239mm}]{geometry}

\crefname{figure}{Fig.}{Figs.}
\Crefname{figure}{Fig.}{Figs.}
\crefname{section}{Sec.}{Secs.}
\Crefname{section}{Sec.}{Secs.}

\DeclareSIUnit\bps{bit/s}
\DeclareSIUnit\Bps{Byte/s}

\usepackage[scaled=0.8]{beramono} 

\lstdefinestyle{mypython}{
    language=Python,
    basicstyle=\ttfamily\scriptsize,  
    keywordstyle=\color{blue!70!black},
    commentstyle=\color{gray},
    stringstyle=\color{orange!80!black},
    numbers=none,
    showstringspaces=false,
    breaklines=false,        
    keepspaces=true,         
    columns=fullflexible,
    aboveskip=2pt,
    belowskip=2pt,
    frame=none               
}

\glsdisablehyper

\newcommand{\ResultNaiveMaxFrequency}{1.42}
\newcommand{\ResultNaiveFrequencyDegradation}{15}

\newcommand{\ResultRoutingOverhead}{76}
\newcommand{\ResultRoutingOverheadPreemptValid}{1}
\newcommand{\ResultRoutingOverheadWideRouter}{78}
\newcommand{\ResultCreditBasedOverhead}{28}
\newcommand{\ResultPreemptValidOverhead}{3}

\newcommand{\ResultMpArea}{270}
\newcommand{\ResultMpMaxFreq}{1.68}
\newcommand{\ResultMpTracks}{2776}

\newcommand{\ResultMpFreqRatio}{-1.5}
\newcommand{\ResultMpTracksRatio}{+76}
\newcommand{\ResultNaiveArea}{275}
\newcommand{\ResultNaiveMaxFreq}{1.42}
\newcommand{\ResultNaiveTracks}{1578}
\newcommand{\ResultNaiveAreaRatio}{+2}
\newcommand{\ResultNaiveFreqRatio}{-16}

\newcommand{\ResultCreditTwoInputArea}{294}
\newcommand{\ResultCreditTwoInputMaxFreq}{1.62}
\newcommand{\ResultCreditTwoInputTracks}{1610}
\newcommand{\ResultCreditTwoInputAreaRatio}{+9}
\newcommand{\ResultCreditTwoInputFreqRatio}{-4.9}
\newcommand{\ResultCreditTwoInputTracksRatio}{+2}
\newcommand{\ResultCreditThreeInputArea}{345}
\newcommand{\ResultCreditThreeInputMaxFreq}{1.62}
\newcommand{\ResultCreditThreeInputTracks}{1610}
\newcommand{\ResultCreditThreeInputAreaRatio}{+28}
\newcommand{\ResultCreditThreeInputFreqRatio}{-4.9}
\newcommand{\ResultCreditThreeInputTracksRatio}{+2}
\newcommand{\ResultPreemptValidArea}{279}
\newcommand{\ResultPreemptValidMaxFreq}{1.70}
\newcommand{\ResultPreemptValidTracks}{1578}
\newcommand{\ResultPreemptValidAreaRatio}{+3}

\makeatletter
\newcommand{\insertnewlines}[1]{%
  \noindent\mbox{}%
  \@tempcnta=#1\relax
  \loop\ifnum\@tempcnta>0
    \\
    \advance\@tempcnta\m@ne
  \repeat
}
\makeatother

\usepackage{circledsteps}
\newcommand\blackcircle[1]{\Circled[inner color=white, outer color=white, fill color=black]{#1}}
\definecolor{orangecirclecolor}{HTML}{F29545}
\newcommand\orangecircle[1]{\Circled[inner color=white, outer color=white, fill color=orangecirclecolor]{#1}}
\definecolor{purplecirclecolor}{HTML}{b12a90}
\newcommand\purplecircle[1]{\Circled[inner color=white, outer color=white, fill color=purplecirclecolor]{#1}}
\definecolor{redcirclecolor}{HTML}{c00000}
\newcommand\redcircle[1]{\Circled[inner color=white, outer color=white, fill color=redcirclecolor]{#1}}
\definecolor{bluecirclecolor}{HTML}{4da4ec}
\newcommand\bluecircle[1]{\Circled[inner color=white, outer color=white, fill color=bluecirclecolor]{#1}}
\definecolor{greencirclecolor}{HTML}{168638}
\newcommand\greencircle[1]{\Circled[inner color=white, outer color=white, fill color=greencirclecolor]{#1}}

\newacronym[plural=WANs, firstplural={Wide Area Networks (WANs)}]{wan}{WAN}{Wide Area Network}
\newacronym[plural=WSNs, firstplural={Wireless Sensor Networks (WSNs)}]{wsn}{WSN}{Wireless Sensor Network}
\newacronym{simd}{SIMD}{Single Instruction Multiple Data}
\newacronym{os}{OS}{Operating System}
\newacronym{ble}{BLE}{Bluetooth Low-Energy}
\newacronym{wifi}{Wi-FI}{Wireless Fidelity}
\newacronym[plural=DVS, firstplural={Dynamic Vision Sensors (DVS)}]{dvs}{DVS}{Dynamic Vision Sensor}
\newacronym{ptz}{PTZ}{Pan-Tilt Unit}

\newacronym[plural=FLLs,firstplural=Frequency Locked Loops (FLLs)]{fll}{FLL}{Frequency Locked Loop}
\newacronym{dram}{DRAM}{Dynamic Random Access Memory}
\newacronym{fpu}{FPU}{Floating Point Unit}
\newacronym{fpss}{FPSS}{Floating Point Subsystem}
\newacronym{frep}{FREP}{Floating Point Repetition}
\newacronym{dma}{DMA}{Direct Memory Access}
\newacronym{ssr}{SSR}{Stream Semantic Register}
\newacronym{issr}{ISSR}{Indirection Stream Semantic Register}
\newacronym[plural=LUTs, firstplural={Lookup Tables (LUTs)}]{lut}{LUT}{Lookup Table}
\newacronym[plural=FPGAs, firstplural={Field Programmable Gate Arrays (FPGAs)}]{fpga}{FPGA}{Field Programmable Gate Array}
\newacronym{dsp}{DSP}{Digital Signal Processing}
\newacronym{mcu}{MCU}{Microcontroller Unit}
\newacronym{spi}{SPI}{Serial Peripheral Interface}
\newacronym{cpi}{CPI}{Camera Parallel Interface}
\newacronym{rf}{RF}{Register File}
\newacronym{fifo}{FIFO}{First-In First-Out Queue}
\newacronym{uart}{UART}{Universal Asynchronous Receiver-Transmitter}
\newacronym{raw}{RAW}{Read After Write}
\newacronym[plural=ISAs, firstplural={Instruction Set Architectures (ISAs)}]{isa}{ISA}{Instruction Set Architecture}
\newacronym{xbar}{XBAR}{crossbar}
\newacronym[firstplural=Scratch-Pad Memories (SPMs)]{spm}{SPM}{Scratch-Pad Memory}
\newacronym{ppa}{PPA}{Power Performance Area}
\newacronym{ipi}{IPI}{Inter-Processor Interrupt}
\newacronym[firstplural=Software-Generated Interrupts (SGIs)]{sgi}{SGI}{Software-Generated Interrupt}
\newacronym{pe}{PE}{Processing Element}
\newacronym{tcdm}{TCDM}{Tightly-Coupled Data Memory}
\newacronym{lsu}{LSU}{Load-Store Unit}
\newacronym{icache}{I\$}{Instruction Cache}
\newacronym{dcache}{D\$}{Data Cache}
\newacronym{wfi}{WFI}{Wait For Interrupt}
\newacronym{gpc}{GPC}{GPU Processing Cluster}
\newacronym{cpu}{CPU}{Central Processing Unit}
\newacronym{gpu}{GPU}{Graphics Processing Unit}
\newacronym[firstplural=General Purpose Graphics Processing Units (GPGPUs)]{gpgpu}{GPGPU}{General Purpose Graphics Processing Unit}
\newacronym{llc}{LLC}{Last-Level Cache}
\newacronym{sm}{SM}{Streaming Multiprocessor}
\newacronym[longplural={Networks-on-Chip}, firstplural={Networks-on-Chip (NoCs)}]{noc}{NoC}{Network-on-Chip}
\newacronym[firstplural=Virtual Channels (VCs)]{vc}{VC}{Virtual Channel}
\newacronym[firstplural=Network Interfaces (NIs)]{ni}{NI}{Network Interface}

\newacronym{dfg}{DFG}{Data Flow Graph}
\newacronym{lcg}{LCG}{Linear Congruential Generator}
\newacronym{prn}{PRN}{Pseudo-Random Number}

\newacronym{ste}{STE}{Straight-Through-Estimator}

\newacronym[plural=PTUs, firstplural={Pan-Tilt Units}]{ptu}{PTU}{Pan-Tilt Unit}
\newacronym{mdf}{MDF}{Medium-density fibreboard}
\newacronym{cvat}{CVAT}{Computer Vision Annotation Tool}
\newacronym{coco}{COCO}{Common Objects in Context}
\newacronym{soa}{SoA}{State of the Art}
\newacronym{sf}{SF}{Sensor Fusion}

\newacronym{dl}{DL}{Deep Learning}
\newacronym{bn}{BN}{Batch Normalization}
\newacronym{FGSM}{FBK}{Fast Gradient Sign Method}
\newacronym{lr}{LR}{Learning Rate}
\newacronym{sgd}{SGD}{Stochastic Gradient Descent}
\newacronym{gd}{GD}{Gradient Descent}
\newacronym{llm}{LLM}{Large Language Model}

\newacronym{sta}{STA}{Static Timing Analysis}

\newacronym[plural=GPIOs, firstplural={General Purpose Inupt Outputs (GPIOs)}]{gpio}{GPIO}{General Purpose Input Output}
\newacronym[plural=LDOs, firstplural={Low Dropout Regulators (LDOs)}]{ldo}{LDO}{Low Dropout Regulator}

\newacronym{inq}{INQ}{Incremental Network Quantization}

\newacronym{CV}{CV}{Computer Vision}
\newacronym{EoT}{EoT}{Expectation over Transformation}
\newacronym{RPN}{RPN}{Region Proposal Network}
\newacronym{TV}{TV}{Total Variation}
\newacronym{NPS}{NPS}{Non-Printability Score}
\newacronym{STN}{STN}{Spatial Transformer Network}
\newacronym{MTCNN}{MTCNN}{Multi-Task Convolutional Neural Network}
\newacronym{YOLO}{YOLO}{You Only Look Once}
\newacronym{SSD}{SSD}{Single Shot Detector}
\newacronym{SOTA}{SOTA}{State of the Art}
\newacronym{NMS}{NMS}{Non-Maximum Suppression}
\newacronym{ic}{IC}{Integrated Circuit}
\newacronym{tcxo}{TCXO}{Temperature Controlled Crystal Oscillator}
\newacronym{jtag}{JTAG}{Joint Test Action Group industry standard}
\newacronym{swd}{SWD}{Serial Wire Debug}
\newacronym{sdio}{SDIO}{Serial Data Input Output}

\newacronym[plural=PCBs, firstplural={Printed Circuit Boards (PCB)}]{pcb}{PCB}{Printed Circuit Board}
\newacronym[plural=ASICs, firstplural={Application Specific Integrated Circuits}]{asic}{ASIC}{Application Specific Integrated Circuit}

\newacronym[plural=BNNs, firstplural={Binary Neural Networks (BNNs)}]{bnn}{BNN}{Binary Neural Network}
\newacronym[plural=NNs, firstplural={Neural Networks}]{nn}{NN}{Neural Network (NNs)}
\newacronym[plural=SCMs, firstplural={Standard Cell Memories (SCMs)}]{scm}{SCM}{Standard Cell Memory}
\newacronym{ann}{ANN}{Artificial Neural Networks}
\newacronym{ml}{ML}{Machine Learning}
\newacronym{ai}{AI}{Artificial Intelligence}
\newacronym{iot}{IoT}{Internet of Things}
\newacronym{fft}{FFT}{Fast Fourier Transform}
\newacronym[plural=OCUs, firstplural={Output Channel Compute Units (OCUs)}]{ocu}{OCU}{Output Channel Compute Unit}
\newacronym{alu}{ALU}{Arithmetic Logic Unit}
\newacronym{mac}{MAC}{Multiply-Accumulate}
\newacronym[longplural={Systems-on-Chip}, firstplural={Systems-on-Chip (SoCs)}]{soc}{SoC}{system-on-chip}
\newacronym[firstplural={multi-processor systems-on-chip (MPSoCs)}]{mpsoc}{MPSoC}{multi-processor system-on-chip}

\newacronym{PGD}{PGD}{Projected Gradient Descend}
\newacronym{CW}{CW}{Carlini-Wagner}
\newacronym{OD}{OD}{Object Detection}

\newacronym{rrf}{RRF}{RADAR Repetition Frequency}
\newacronym{nlp}{NLP}{Natural Language Processing}
\newacronym{qam}{QAM}{Quadrature Amplitude Modulation}
\newacronym{rri}{RRI}{RADAR Repetition Interval}
\newacronym{radar}{RADAR}{Radio Detection and Ranging}
\newacronym{loocv}{LOOCV}{Leave-one-out cross validation}

\newacronym{bsp}{BSP}{Board Support Package}
\newacronym{ttn}{TTN}{The Things Network}
\newacronym{wip}{WIP}{Work in Progress}
\newacronym{json}{JSON}{JavaScript Object Notation}
\newacronym{qat}{QAT}{Quantization-Aware Training}

\newacronym{cls}{CLS}{Classification Error}
\newacronym{loc}{LOC}{Localization Error}
\newacronym{bkgd}{BKGD}{Background Error}
\newacronym{roc}{ROC}{Receiver Operating Characteristic}
\newacronym{frr}{FRR}{False Rejection Rate}
\newacronym{eer}{EER}{Equal Error Rate}
\newacronym{snr}{SNR}{Signal-to-Noise Ratio}
\newacronym{flop}{FLOP}{Floating-Point Operation}
\newacronym{fp}{FP}{Floating-Point}
\newacronym{fps}{FPS}{Frames Per Second}
\newacronym{oi}{OI}{Operational Intensity}
\newacronym[first={IPC (Instructions per Cycle)}]{ipc}{IPC}{Instructions per Cycle}

\newacronym{gsc}{GSC}{Google Speech Commands}
\newacronym{mswc}{MSWC}{Multilingual Spoken Words Corpus}
\newacronym{demand}{DEMAND}{Diverse Environments Multichannel Acoustic Noise Database}

\newacronym[plural=SNNs, firstplural={Spiking Neural Networks (SNNs)}]{snn}{SNN}{Spiking Neural Network}
\newacronym[plural=DNNs, firstplural={Deep Neural Networks (DNNs)}]{dnn}{DNN}{Deep Neural Network}
\newacronym[plural=TCNs,firstplural=Temporal Convolutional Networks]{tcn}{TCN}{Temporal Convolutional Network}
\newacronym[plural=CNNs,firstplural=Convolutional Neural Networks (CNNs)]{cnn}{CNN}{Convolutional Neural Network}
\newacronym[plural=TNNs,firstplural=Ternarized Neural Networks]{tnn}{TNN}{Ternarized Neural Network}
\newacronym{ds-cnn}{DS-CNN}{Depthwise Separable Convolutional Neural Network}
\newacronym{rnn}{RNN}{Recurrent Neural Network}
\newacronym{gcn}{GCN}{Graph Convolutional Network}
\newacronym{mhsa}{MHSA}{Multi-Head Self Attention}
\newacronym{crnn}{CRNN}{Convolutional Recurrent Neural Network}
\newacronym{clca}{CLCA}{Convolutional Linear Cross-Attention}

\newacronym{bf}{BF}{Beamforming}
\newacronym{anc}{ANC}{Active Noise Cancellation}
\newacronym{agc}{AGC}{Automatic Gain Control}
\newacronym{se}{SE}{Speech Enhancement}
\newacronym{mct}{MCT}{Multi-Condition Training}
\newacronym{mcta}{MCTA}{Multi-Condition Training \& Adaptation}
\newacronym{pcen}{PCEN}{Per-Channel Energy Normalization}
\newacronym{mfcc}{MFCC}{Mel-Frequency Cepstral Coefficient}
\newacronym{asr}{ASR}{Automated Speech Recognition}
\newacronym{kws}{KWS}{Keyword Spotting}
\newacronym{odl}{ODL}{On-Device Learning}

\newacronym{nl-kws}{NL-KWS}{Noiseless Keyword Spotting}
\newacronym{na-kws}{NA-KWS}{Noise-Aware Keyword Spotting}
\newacronym{odda}{ODDA}{On-Device Domain Adaptation}
\newacronym{hpm}{HPM}{High-Performance Mode}
\newacronym{lpm}{LPM}{Low-Power Mode}
\newacronym{ge}{gate equivalents}{GEs}

\def\BibTeX{{\rm B\kern-.05em{\sc i\kern-.025em b}\kern-.08em
    T\kern-.1667em\lower.7ex\hbox{E}\kern-.125emX}}
\begin{document}

\title{Physically-Aware Preemptive Virtual Channels for Deadlock-Free AXI Networks-on-Chip}

\ifdefined\blindreview
\author{\centering{\textit{Authors omitted for blind review.}}}
\else
\author{
    \IEEEauthorblockN{Lorenzo Leone\textsuperscript{1}\orcidlink{0009-0000-3976-847X}, Luca Colagrande\textsuperscript{1}\orcidlink{0000-0002-7986-1975}, Luca Benini\textsuperscript{1,2}\orcidlink{0000-0001-8068-3806}}
    \IEEEauthorblockA{\textsuperscript{1}\textit{ETH Zürich, Zürich, Switzerland}, \textsuperscript{2}\textit{Università di Bologna, Bologna, Italy}}
    \{lleone, colluca, lbenini\}@iis.ee.ethz.ch
}
\fi

\maketitle

\begin{abstract}
    As many-core \glspl{soc} continue to scale, \glspl{noc} must sustain increasingly high memory bandwidth while preserving deadlock freedom.
    In AXI4 systems, protocol-level dependencies between read and write traffic can create circular waits at the network endpoints, even when the routing algorithm itself is deadlock-free.
    Decoupling these traffic classes avoids such dependencies, but exposes a key implementation trade-off: multiplane \glspl{noc} duplicate wide physical links and increase routing pressure, whereas conventional \gls{vc} routers add substantial control complexity, area, and timing overhead.
    This work revisits this trade-off for modern wide-link \glspl{noc}. We evaluate four deadlock-free AXI4 traffic-class separation schemes: a multiplane baseline and three lightweight \gls{vc}-based designs.
    Among these designs, we propose Preemptive \glspl{vc}, a physically-aware architecture that can save up to \ResultRoutingOverhead\% of link resources with comparable frequency and only \ResultPreemptValidOverhead\% router area overhead relative to the multiplane design.
\end{abstract}


\section{Introduction}
\label{sec:introduction}
As many-core SoCs continue to scale \cite{noc-survey}, the memory traffic generated by large numbers of processing elements increases sharply, making high-bandwidth and scalable \glspl{noc} essential for sustaining performance.
At the same time, providing industry-standard AXI4 \cite{arm_axi_spec} interfaces at the network boundaries remains critical for seamless integration of existing IPs, usually designed with AXI4 initiator and/or target interfaces \cite{nightingale2024, nvdla_hw_spec, amd_versal_noc}.

\begin{figure}[t]
    \centering
    \includegraphics[width=1\columnwidth]{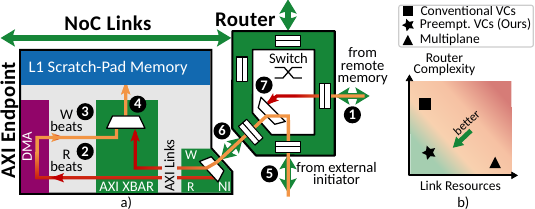}
    \begingroup
        \phantomsubcaption
        \label{fig:deadlock}
        \phantomsubcaption
        \label{fig:challenges}
    \endgroup
    \vspace{-1em}
    \caption{a) Deadlock scenario: all links are blocked (red arrowheads) or starved (red tails); b) Router design space.}
    \vspace{-0.5em}
\end{figure}

Deadlock freedom remains a central requirement in \glspl{noc}, both at the routing and protocol levels \cite{coffman1971,hansson2007,concer2009}.
Endpoint behavior in AXI4-based systems creates protocol dependencies that are invisible to the routing algorithm itself.
Thus, even when deterministic XY routing is employed, protocol-level deadlocks may still arise if AXI4-capable endpoints internally couple read and write transactions \cite{hansson2007}.
\Cref{fig:deadlock} illustrates a representative case in which a DMA engine (AXI4 initiator) issues a read burst request to a remote memory \blackcircle{1}.
As the read data returns, each AXI4 beat \blackcircle{2} is immediately forwarded into a write burst \blackcircle{3} directed to a local L1 \gls{spm} (AXI4 target) through the AXI4 crossbar \blackcircle{4}.
If, before the DMA completes, an external initiator issues a write burst to the same L1 \gls{spm} \blackcircle{5}, its request may occupy the local \gls{noc} link \blackcircle{6}.
The DMA can then no longer make forward progress on the read-response path \blackcircle{7}, and a protocol-level \emph{circular wait} arises at the network endpoint, stalling the entire system.
More generally, the challenge of protocol- or message-dependent deadlocks caused by the interaction between \glspl{noc} and endpoint protocols has long been recognized in the literature \cite{hansson2007,concer2009,li2024vn}. To eliminate protocol-level deadlocks without constraining endpoint behavior, AXI4 read and write channels must also be decoupled so that they cannot block one another on the same \gls{noc} link.

One solution is to use \glspl{vc} \cite{dally1987, dally1992}.
As shown in \Cref{fig:old-router}, \glspl{vc} provide separate buffers for different data streams while preserving a shared physical link.
In our case, AXI4 read and write transactions can be assigned to distinct buffers, so that returning read data \blackcircle{7} and incoming write data \blackcircle{5} no longer compete for the same downstream path in the local interconnect.
A flit may occupy the shared link only if the corresponding downstream buffer has free space, ensuring that the link is released in the next cycle.
A second solution is the multiplane approach \cite{Dally2022-keynote,carloni-mpvc}, which assigns different traffic classes to separate physical links, thereby avoiding inter-class blocking on shared links.

Prior work has established the conventional view that \glspl{vc} are more complex than multiplane designs \cite{carloni-mpvc,mp-with-vc,mullins2004}, largely because earlier studies considered classical \gls{vc} routers targeted at improving link utilization, which incur significant control overhead.
Under this assumption, paying for additional link resources has often been seen as preferable to the area and timing cost of full \gls{vc} support.

However, the design space of modern \glspl{noc} is shifting, fundamentally changing the trade-off between router complexity and link resources (\Cref{fig:challenges}).
As accelerator-centric systems continue to scale, integrating hundreds of thousands of AI cores, on-chip networks must sustain aggregate bandwidth demands reaching tens of \si{\peta\Bps} \cite{cerebras23,cerebras24,waferLlm25,zhang2025}.
To sustain this bandwidth, \gls{noc} link widths have therefore increased, both in proprietary, closed-source solutions \cite{davinci19, dojo22, blackhole24, arteris_xl_2xl} and in open-source designs such as FlooNoC~\cite{floonoc}.
The latter stands out as the first open-source AXI4-compliant \gls{noc}, supporting high-throughput transfers with links up to 1024 bits \cite{flatattention}.
At such link widths, the routing-resource cost of physical-plane duplication becomes critical.
Moreover, as these fabrics scale toward wafer-scale integration, reliability also becomes increasingly important.
Achieving reliability often relies on redundancy, with examples of triple-redundant interconnects incurring up to a 3$\times$ increase in link width, further exacerbating wiring demand and routing pressure \cite{cerebraspatent22,amd_pg406_noc2_2025,axilite24,ouyang24,chang20}.

Under these stringent constraints, \gls{vc} implementations re-emerge as a promising option to preserve protocol-level deadlock freedom while saving valuable routing resources.
To this end, we propose \textit{Preemptive \glspl{vc}}, a novel lightweight \gls{vc} router design for deadlock-free AXI4 traffic separation which preserves the resource parsimony of \glspl{vc} while approaching the area and timing characteristics of multiplane router designs.

To summarize our contributions, we:
\begin{itemize}
    \item Extend an open-source AXI4 NoC design (FlooNoC \cite{floonoc}) with four deadlock-free AXI4 traffic-class separation schemes, including a multiplane design and three lightweight \gls{vc}-based designs. Among these, we propose a novel design, \textit{Preemptive \glspl{vc}}, which balances timing, area, routing resources and bandwidth utilization%
    \footnote{ Our implementations can be found at:
\ifdefined\blindreview
    https://hidden-for-double-blind-review.com
\else
    \url{https://github.com/pulp-platform/FlooNoC/releases/tag/v0.8.0}
\fi
}%
.
    \item Integrate the proposed designs into a representative mesh-based \gls{soc} to demonstrate their applicability in a realistic AXI4-capable tiled system%
    \footnote{
\ifdefined\blindreview
    https://hidden-for-double-blind-review.com
\else
    \url{https://github.com/Lore0599/gwaihir/tree/vc-exploration}
\fi
}%
.
    \item Evaluate timing, area, routing resource usage and bandwidth utilization for all designs in \textsc{TSMC} \SI{7}{\nano\meter}, showing that Preemptive \glspl{vc} can save \ResultRoutingOverhead\% of routing resources while maintaining comparable frequency and incurring only \ResultPreemptValidOverhead\% area overhead relative to the multiplane solution.
\end{itemize}

\section{Related Work}
\begin{figure}[t]
    \centering
    \includegraphics[width=0.92\columnwidth]{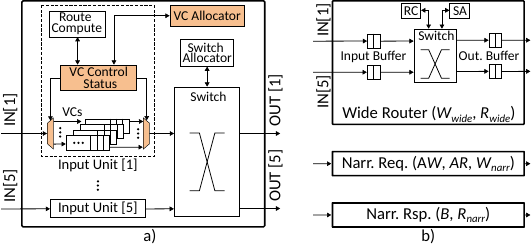}
    \begingroup
        \phantomsubcaption
        \label{fig:old-router}
        \phantomsubcaption
        \label{fig:floonoc}
    \endgroup
    \caption{a) Conventional four-stage \gls{vc}-based router; b) baseline multiplane FlooNoC single-cycle router.}
    \label{fig:router-comparison}
    \vspace{-0.5em}
\end{figure}

Originally introduced to break cyclic channel dependencies and guarantee deadlock-free routing \cite{dally1987}, \glspl{vc} were soon adopted to improve link utilization by allowing multiple logical flows to share the same physical link \cite{dally1992}.
This evolution led to the conventional four-stage router microarchitecture (\Cref{fig:old-router}), designed to dynamically map incoming packets to the available \glspl{vc}.
In this architecture, routing computation (\texttt{RC}) and switch allocation (\texttt{SA}) must be complemented by substantial \gls{vc} control and arbitration logic to support dynamic allocation (\texttt{VA}).
Consequently, interest in reduced-complexity \gls{vc} designs has resurfaced to mitigate this overhead.
Xu et al. \cite{xu2010} simplify \gls{vc} and switch arbitration through partially constrained allocation schemes, but still retain a dynamic allocator to adapt \glspl{vc} to traffic fluctuations.
Shim et al.~\cite{shim2009} statically bind \glspl{vc} to source--destination traffic paths encoded in routing tables, targeting routing optimization rather than traffic-class separation.
In contrast, our work targets protocol-driven AXI4 traffic-class decoupling: by statically associating each logical channel with one \gls{vc}, much of the allocation, control, and arbitration complexity present in conventional and prior reduced-complexity \gls{vc} routers can be removed, enabling lightweight \glspl{vc} for deadlock-free traffic-class separation..

Multiple physical networks, or multiplanes, have long been used as an alternative to \glspl{vc} to separate interdependent traffic classes and avoid protocol-level deadlock.
Hansson et al. \cite{hansson2007} note that many \glspl{noc} eliminate request--response dependencies by introducing separate physical networks.
The trade-off between \glspl{vc} and multiplanes was previously studied by Yoon et al. \cite{carloni-mpvc}, who showed that multiplanes can achieve simpler routers and higher frequency than classical \gls{vc} routers, at the cost of fabric duplication.
However, while their study targets conventional dynamically allocated \gls{vc} routers, our work revisits this trade-off under different design constraints, where FlooNoC's lightweight single-cycle router (\Cref{fig:floonoc}) and the static use of \glspl{vc} for AXI4 traffic-class separation fundamentally change the cost--performance balance.

Following the multiplane approach, other works have adopted multiple physical networks for traffic-class separation.
In the \gls{gpgpu} domain, designs such as RAPID \cite{rapid2018} and DA2 \cite{hanjoon2012} use two separate 128-bit networks to exploit the request--reply traffic asymmetry of \gls{gpgpu} memory systems.
Similarly, tiled manycore systems such as OpenPiton \cite{piton2018} and ESP \cite{tombesi2023,santos2024} rely on multiple physical networks, using, respectively, three and six 64-bit physical planes to provide bandwidth and avoid deadlock.
While multiplanes have been adopted successfully for message-class separation, they have mostly been demonstrated on relatively narrow links, which are not suited for modern large-scale, data-intensive AI-oriented \glspl{soc}. In such systems, wide links up to 1024 bits are common to sustain high-bandwidth data movement \cite{davinci19, dojo22, blackhole24, arteris_xl_2xl}.

FlooNoC \cite{floonoc} stands out as the first open-source AXI4-compliant \gls{noc} targeting AI accelerators with very wide links.
As shown in \Cref{fig:floonoc}, it implements two logical networks, one wide (up to 1024 bits) and one narrow (typically 64 bits), on top of three shared physical planes: one narrow channel for AXI4 requests (narrow and wide \texttt{AW} and \texttt{AR}, and narrow \texttt{W}), one narrow channel for AXI4 responses (narrow and wide \texttt{B}, and narrow \texttt{R}), and one wide channel for AXI4 data (wide \texttt{W} and \texttt{R}).
This organization makes FlooNoC's original multiplane choice inexpensive: only narrow traffic is physically separated, while the costly wide data plane remains shared between AXI4 read and write data streams.
In the deadlock scenario of \Cref{sec:introduction}, however, the wide AXI4 read and write data streams must also be separated.
A pure multiplane solution would therefore duplicate the wide tile-to-tile link, sharply increasing routing pressure.

\section{Architecture}
\label{sec:architecture}
\begin{figure}[t]
    \centering
    \includegraphics[width=\columnwidth]{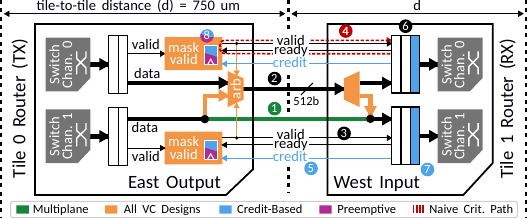}
    \caption{Wide tile-to-tile connection for the various designs.}
    \label{fig:architecture}
\end{figure}

The deadlock scenario in \Cref{fig:deadlock} stems from the interaction between wormhole-routed AXI4 transactions and IP blocks, such as DMAs, that internally couple read and write streams.
This coupling can create circular dependencies and lead to protocol-level deadlocks, which can be avoided by breaking at least one of Coffman's conditions \cite{coffman1971}.
We study the design space through a physically-aware lens, since the router-complexity vs. link-resource trade-off discussed in \Cref{sec:introduction} is critical to system-level feasibility.
We evaluate four architectural solutions (\Cref{fig:architecture}), each targeting a different deadlock condition.
We focus on wide data routers and the respective tile-to-tile links, which represent the main bottleneck for decoupling AXI4 read and write data streams in large accelerator-centric systems.

\subsubsection{Multiplane}
The baseline solution extends the multiplane approach adopted in FlooNoC.
While FlooNoC already separates narrow AXI4 request and response traffic, it uses a single shared wide plane for both read and write channels.
To decouple these traffic classes, we instantiate two independent wide planes, one for read and one for write data, each with its own tile-to-tile physical link \greencircle{1} and router datapath.
As a result, read and write data streams no longer contend for the same wide physical channel, breaking Coffman's \emph{mutual exclusion} condition.
The main cost is the duplication of the wide tile-to-tile link, which doubles the number of wide pins and routed tracks crossing tile boundaries.
Since the wide link accounts for the majority (\ResultRoutingOverheadWideRouter\%) of the tile-to-tile link resources \cite{floonoc}, the multiplane solution substantially increases routing pressure.

\subsubsection{Naive}
To reduce routing overhead, an alternative approach is to use \glspl{vc}.
We first evaluate a \emph{Naive} valid/ready implementation \cite{dally1986torus}, where a single physical data link is shared across \glspl{vc} \blackcircle{2}, while the \texttt{valid} and \texttt{ready} handshake signals are replicated per \gls{vc} \blackcircle{3}%
\footnote{As in the multiplane design, our implementation also replicates the switches. This could be further optimized by sharing one switch between \glspl{vc}, as shown in \Cref{fig:old-router}.}.
This allows the sender to select the active stream and the receiver to apply independent backpressure for each traffic class.
To break Coffman's \emph{hold and wait} condition, data is injected onto the shared physical link only when the corresponding downstream receiver is ready, by masking the upstream \texttt{valid} with the downstream \texttt{ready}.
However, this introduces a \texttt{ready}-to-\texttt{valid} combinational dependency, creating a logic path from the receiving router, through the transmitting router, and back to the receiving router \redcircle{4}.
In tile-based architectures, this path spans twice the tile width (\texttt{d}), making physical delay a critical timing contributor and significantly limiting the achievable frequency.

\begin{figure}[t]
    \centering
    \includegraphics[width=\columnwidth]{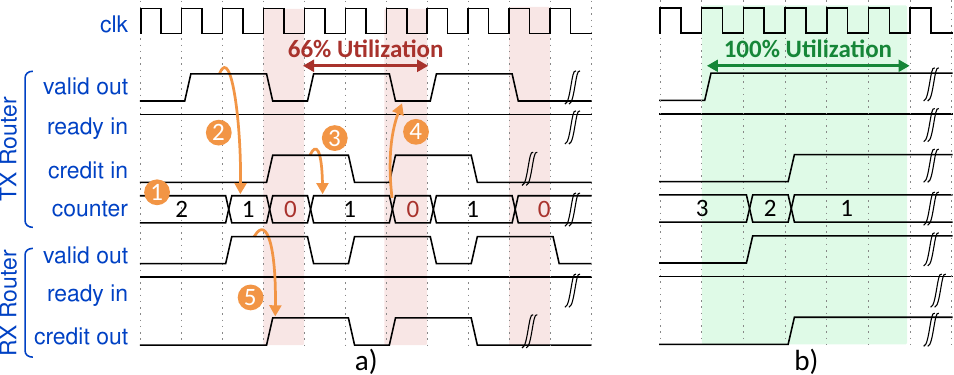}
    \caption{\texttt{CreditBased} timing with FIFO depth of a) two and b) three flits.}
    \begingroup
        \phantomsubcaption
        \label{fig:credit-util-bad}
        \phantomsubcaption
        \label{fig:credit-util-good}
    \endgroup
    \label{fig:credit-util}
    \vspace{-0.5em}
\end{figure}

\subsubsection{CreditBased}
To remove the \texttt{ready}-to-\texttt{valid} combinational dependency, a CreditBased protocol can be adopted by extending the router interface with a credit signal \bluecircle{5} that indicates flit consumption in the downstream input buffer \blackcircle{6} \cite{dally-book}.
Each output \gls{vc} has a credit counter \bluecircle{8}, initialized to the downstream FIFO depth \orangecircle{1} (\Cref{fig:credit-util}), decremented when a flit is transmitted \orangecircle{2}, and incremented when the downstream router returns a credit after the corresponding FIFO entry is freed \orangecircle{3}.
A flit is injected onto the link only if the corresponding counter is non-zero \orangecircle{4}, guaranteeing downstream buffer space and breaking Coffman's \emph{hold and wait} condition.

Compared to the \texttt{Naive} solution, this scheme removes the long combinational backpressure path by registering the downstream credit.
However, the returned credit is visible upstream only one cycle after the downstream flit consumption \orangecircle{5}, delaying the counter increment.
As shown in \Cref{fig:credit-util}, this credit delay can reduce throughput by up to 33\%.
To hide it and preserve full bandwidth, the downstream FIFO must absorb one additional flit, implying a minimum depth of three flits \bluecircle{7}.

\subsubsection{Preemptive}
To the best of our knowledge, prior \gls{vc} designs have not targeted Coffman's \emph{no preemption} condition.
We therefore propose a novel \texttt{Preemptive} scheme: any valid input may acquire the shared physical link, but can be preempted by another valid stream if its downstream receiver is not ready.
Link ownership is selected through round-robin arbitration among the valid streams, while the downstream \texttt{ready} signals are used to determine whether the currently selected stream may retain the link in the following cycle.

Compared to the \texttt{Naive} solution, this approach restores the target frequency by removing the long \texttt{ready}-to-\texttt{valid} combinational path.
This is achieved by registering the downstream \texttt{ready} \purplecircle{8}, preventing it from driving \texttt{valid} combinationally: \texttt{ready} only affects next-cycle link ownership, allowing a stalled stream (\texttt{ready == 0}) to be preempted in the next cycle.
At the same time, unlike the \texttt{CreditBased} design, the \texttt{Preemptive} solution does not need additional input buffering to sustain full bandwidth.
The sender does not wait for delayed credit availability before driving the link; instead, it speculatively asserts \texttt{valid} for the selected stream.
If that stream cannot make progress (\texttt{ready == 0}), its flit remains buffered at the source side and another valid stream with an available downstream receiver can take over the shared link in the next cycle.
Thus, the feedback-latency window that the \texttt{CreditBased} scheme must hide with a deeper downstream FIFO is avoided.

\section{Results}
\label{sec:results}

\begin{figure}[t]
    \centering
    \includegraphics[width=\columnwidth]{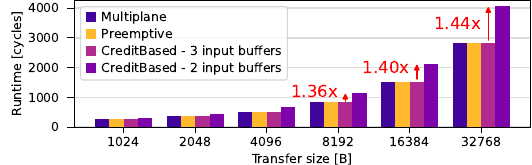}
    \caption{Runtime of a 2D broadcast transfer to all 16 tiles.}
    \label{fig:dma-transfers}
    \vspace{-0.5em}
\end{figure}

To evaluate the proposed designs in a realistic AXI4-capable tiled system, we integrate them into a $4 \times 4$ mesh \gls{soc}, where each tile contains a Snitch cluster \cite{zaruba2021} connected to a FlooNoC network extended with the deadlock-free wide-router variants.

Performance is evaluated through cycle-accurate simulation in \textsc{QuestaSim 2023.4}, while physical implementation is carried out in \textsc{TSMC} \SI{7}{\nano\meter} technology using \textsc{Fusion Compiler 2024.09} under worst-case conditions (SS, $-40^\circ$C, \SI{0.675}{\volt}).

\subsection{Performance evaluation}
To assess the impact of the four router designs, we consider a representative use case in which data are broadcast in burst mode to all 16 tiles using a binary-tree algorithm \cite{sanders2019}.
\Cref{fig:dma-transfers} shows the runtime for different transfer sizes.
To highlight the impact of insufficient buffering in the \texttt{CreditBased} design, we also include a configuration with only two input buffers.

All designs except for the two-buffer \texttt{CreditBased} configuration achieve the same performance as the multiplane solution, showing that lightweight \glspl{vc} do not introduce runtime degradation.
As discussed in \Cref{sec:architecture}, the two-buffer \texttt{CreditBased} design suffers from a 33\% throughput loss, translating into an asymptotic 1.5$\times$ increase in runtime (3 instead of 2 cycles for every 2 data beats).
For small transfers, the reduced link utilization is partially hidden by fixed costs, such as transfer setup, round-trip latency, and tile synchronization, while for larger sizes the runtime overhead reaches the expected 1.5$\times$.

\subsection{Area and timing evaluation}
In large-scale systems, tile-to-tile wire delay has a major impact on the achievable frequency, since \gls{noc} links typically span the full tile width and may extend even further in low-diameter \gls{noc} topologies with bypass connections \cite{yanghui2020}.
At the same time, the exact tile dimensions depend on the target system, and the corresponding tile-to-tile link length may vary accordingly.
Moreover, performing place-and-route on the complete cluster tile would make the critical path strongly dependent on the internal compute logic of the cluster, such as the FPU, caches, or other local blocks, thereby obscuring the actual timing impact of the proposed router implementations.
For this reason, we place and route two FlooNoC routers at a representative tile-to-tile distance of \SI{750}{\micro\meter} \cite{floonoc}, capturing inter-router wire delay and excluding unrelated cluster logic.

\begin{figure}[t]
    \centering
    \begin{minipage}[t]{\columnwidth}
        \centering
        \includegraphics[width=\textwidth]{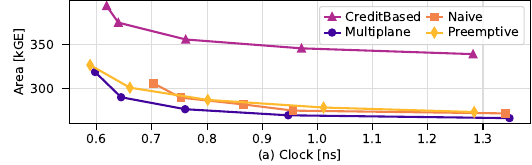}
    \end{minipage}

    \begin{minipage}[t]{\columnwidth}
        \centering
        \begin{minipage}[t]{0.65\textwidth}
            \centering
            \includegraphics[width=\textwidth]{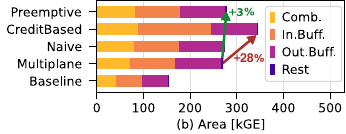}
        \end{minipage}
        \hfill
        \begin{minipage}[t]{0.33\textwidth}
            \includegraphics[width=\textwidth]{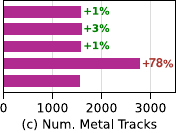}
        \end{minipage}
        \end{minipage}
    \vspace{-1em}
    \begingroup
        \phantomsubcaption
        \label{fig:at_plot}
        \phantomsubcaption
        \label{fig:area-breakdown}
        \phantomsubcaption
        \label{fig:routing-resources}
    \endgroup
    \caption{a) AT plots of the analyzed deadlock-free designs.
             b) Area breakdown at \SI{1.1}{GHz}.
             c) Routing usage, \%w.r.t. baseline.}
    \label{fig:results}
    \vspace{1.5em}
\end{figure}

\begin{table}[t]
    \centering
    \caption{}
    \label{tab:results}
    \resizebox{\columnwidth}{!}{%
    \begin{tabular}{lclll}
        \hline
        Configuration & BW util. & Area [kGE] & Max. Freq.\ [GHz] & \#Metal tracks \\
        \hline
        \texttt{Multiplane}             & 100\% & \textbf{\ResultMpArea}             & \ResultMpMaxFreq~(\ResultMpFreqRatio\%)             & \ResultMpTracks~(\ResultMpTracksRatio\%)             \\
        \texttt{Naive}                  & 100\% & \ResultNaiveArea~(\ResultNaiveAreaRatio\%)          & \ResultNaiveMaxFreq~(\ResultNaiveFreqRatio\%)          & \ResultNaiveTracks          \\
        \texttt{CreditBased} 2 buff.      &  66\% & \ResultCreditTwoInputArea~(\ResultCreditTwoInputAreaRatio\%) & \ResultCreditTwoInputMaxFreq~(\ResultCreditTwoInputFreqRatio\%) & \ResultCreditTwoInputTracks~(\ResultCreditTwoInputTracksRatio\%) \\
        \texttt{CreditBased} 3 buff.      & 100\% & \ResultCreditThreeInputArea~(\ResultCreditThreeInputAreaRatio\%) & \ResultCreditThreeInputMaxFreq~(\ResultCreditThreeInputFreqRatio\%) & \ResultCreditThreeInputTracks~(\ResultCreditThreeInputTracksRatio\%) \\
        \texttt{Preemptive}          & \textbf{100\%} & \ResultPreemptValidArea~(\ResultPreemptValidAreaRatio\%)   & \textbf{\ResultPreemptValidMaxFreq}   & \textbf{\ResultPreemptValidTracks}   \\
        \hline
        \multicolumn{5}{l}{\footnotesize Percentages are relative to the best value for each metric, highlighted in bold.} \\
    \end{tabular}}
    \vspace{-1.5em}
\end{table}
We start by assessing the \texttt{Multiplane} solution relative to the FlooNoC baseline \cite{floonoc}.
Adding a second 512-bit router doubles the area (\Cref{fig:area-breakdown}) while preserving the achievable frequency.
However, it increases routing-resource usage by \ResultRoutingOverheadWideRouter\% (\Cref{fig:routing-resources}), significantly increasing routing complexity.

The \texttt{Naive} \gls{vc} implementation matches the baseline in routing cost, but the \texttt{ready}-to-\texttt{valid} dependency extends the critical path, lowering the achievable frequency to \SI{\ResultNaiveMaxFrequency}{\giga\hertz}, a \ResultNaiveFrequencyDegradation\% reduction relative to the \texttt{Multiplane} solution (\Cref{fig:at_plot}).
This result highlights the strong impact of tile-to-tile interconnect delay in realistic tile-based systems.

The \texttt{CreditBased} design removes this dependency and restores timing to the level of the \texttt{Multiplane} approach.
However, the additional input buffer required to sustain full throughput increases area by \ResultCreditBasedOverhead\%, highlighting the high cost of the extra buffering demanded by this solution.

Finally, the \texttt{Preemptive} approach keeps the router area close to that of the \texttt{Multiplane} design, with a negligible $\ResultPreemptValidOverhead\%$ overhead coming from the lightweight arbitration and masking logic.
Compared to the \texttt{Naive} \gls{vc} design, it restores the achievable frequency to the level of the \texttt{Multiplane} approach.
At the same time, unlike the \texttt{Multiplane} solution, the \texttt{Preemptive} approach increases routing-resource usage by only \ResultRoutingOverheadPreemptValid\% over the baseline, due solely to the additional \texttt{valid} and \texttt{ready} signals required for the two virtual channels.

\section{Conclusion}
In this work, we presented a detailed analysis of architectural solutions to prevent protocol-level deadlocks in \glspl{noc} with AXI4 interfaces (\Cref{tab:results}).
We proposed Preemptive \glspl{vc}, a novel lightweight \gls{vc} implementation that incurs no frequency degradation relative to multiplane designs, with a modest router area overhead of just \ResultPreemptValidOverhead\%, while saving up to \ResultRoutingOverhead\% of link routing resources, a precious commodity in modern \glspl{noc} with wide, high-bandwidth physical links.
\bibliography{paper}
\bibliographystyle{IEEEtran}

\end{document}